\documentclass[aps,fleqn,showpacs,twocolumn]{revtex4}

\usepackage{graphicx}
\usepackage{amsmath} 

\def\Fig#1{Figure~\ref{#1}}
\def\fig#1{Fig.\,\ref{#1}}
\def\eq#1{Eq.\,(\ref{#1})}
\def\EF{F}
\def\coll#1{[#1]}

\def\prhoi{p'_{\rho}}\def\d{\mathrm d}\def\phiV{\phi_{{}_{V}}}

\def\zc{z^{\star}}\def\pc{p^{\star}_{z}}\def\rhoc{\rho^{\star}}
\def\phic{\phi^{\star}}\def\xc{x^{\star}}
\def\cc{c^{\star}}\def\Cc{C^{\star}}
\def\zquiv{\tilde{z}}\def\xquiv{\tilde{x}}

\begin{document}
\title{Electron-energy bunching in laser-driven soft recollisions}
\author{Alexander K\"astner$^{1}$, Ulf Saalmann$^{1,2}$, and Jan M.\ Rost$^{1,2}$}
\address{$^1$Max Planck Institute for the Physics of Complex Systems,
   N\"othnitzer Stra{\ss}e 38, 01187 Dresden, Germany\\
   $^2$Max Planck Advanced Study Group at CFEL,
   Luruper Chaussee 149, 22761 Hamburg, Germany} 

\date{\today}

\begin{abstract}\noindent 
We introduce soft recollisions in laser-matter interaction. They are characterized by the electron missing the ion upon recollision 
in contrast to the well-known head-on collisions responsible for high-harmonic generation or above-threshold ionization. We demonstrate analytically that soft recollisions can cause a bunching of photo-electron energies through which a series of low-energy peaks emerges in the electron yield along the laser polarization axis. This peak sequence is universal, it does not depend on the binding potential, and is found below an excess energy of one fifth of the ponderomotive energy.
\end{abstract}
\maketitle

\noindent 
Recollision of an electron with its parent ion under a linearly polarized strong laser field has been shown to be the basis of a plethora of phenomena in atoms
\cite{co93,ku87}, molecules \cite{itle+04}, clusters \cite{saro08}
and solids \cite{br87}. In principle the recollision process is very
simple and a single degree of freedom along the laser
polarization axis is sufficient to describe it (often referred to as the
three-step model \cite{co93}): Firstly, the bound
electron is released from an atom due to the strong electric field of
a laser. Secondly, it is accelerated and driven back to the ion. In
the third step it either recombines in the atomic potential or is scattered from it. In the former case, high-order harmonics are generated (HHG) due to recombination of the electron \cite{leba+94}. In the latter case, the elastic head-on collision induces the high-energy phenomenon
 of above-threshold ionization
(ATI) with fast electrons emitted \cite{pabe+94,mipa+06}. 
The enormous impact of HHG up to recent proposals for imaging of molecular orbitals \cite{itle+04} and the generation of attosecond pulses \cite{kriv09} is not the least due to the simple yet accurate description with the three-step model.

Recently, a surprising strong peak\,---\,the ``low-energy structure'' (LES)\,---\,was observed at few eV in the photo-electron spectrum of atoms in strong infra-red (a few $\mu$m wavelength) laser pulses \cite{blca+09,quli+09} and confirmed numerically with classical calculations \cite{liha10,yapo+10}. Although the LES peak contains about half of the photo electrons it was not seen in any of the numerous experiments done with 800\,nm laser pulses.

Here, we will give an analytical explanation of the LES by introducing a low-energy soft-recollision mechanism. It gives rise to a universal series of low-energy peaks in the momentum spectrum of the photo electron with well defined relative positions of 3/5, 5/7, 7/9 \ldots\ on an absolute energy scale of about one fifth of the ponderomotive energy $F^{2}/(4\omega^{2})$, where $F$ is the amplitude and $\omega$ the frequency of the laser field. These peaks do not require a special binding potential, e.\,g., long range, nor do they need more than one degree of freedom to appear, and they can be derived classically since they rely essentially on the well known strong-field trajectories as will become clear later.

\begin{figure*}[t]
\begin{center}
\includegraphics[width=\textwidth]{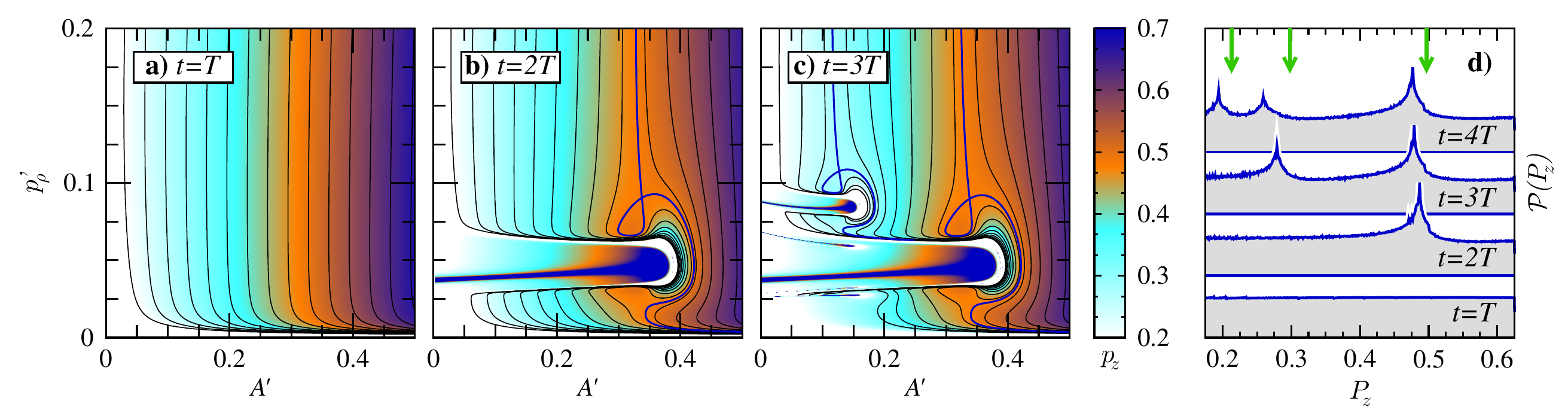}
\caption{(Color online) The deflection function $p_{z}$ for the momentum along the laser axis as a function of the initial vector potential $A'{=}A\sin\phi'$ and the initial transversal momentum $\prhoi$ shown after a) one, b) two, and c) three laser
cycles. Spectra according to \eq{eq:spectrum} with $w=1$ integrated over the initial parameter range of panels a--c are shown in the right panel d).
The arrows point to the predicted momenta according to \eq{eq:peaks}. The laser has a wavelength of
$\lambda{=}2\,\mu$m ($\omega{=}0.0228$\,a.u.) at an intensity of
$I{=}10^{14}$W/cm$^{2}$ ($F{=}0.0534$\,a.u.).
}
\label{fig:contours123}
\end{center}
\end{figure*}%
We will begin by working out the classical structures which are responsible for the LES \cite{blca+09}, i.e., we consider a Hamiltonian $H = H_{0}+V$ with (throughout the paper we use atomic units unless stated otherwise)
\begin{subequations}\label{eq:ham}\begin{align}
\label{eq:ham1}
H_{0}&=\mathbf{p}^{2}/2+z\,\EF \cos(\omega t)
\\ 
\label{eq:ham2}
V&=-1/\left(\rho^{2}{+}z^{2}\right)^{1/2},
\end{align}\end{subequations}
describing an electron with position $\mathbf{r}{\equiv}(\rho,z)$ and momentum $\mathbf{p}{\equiv}(p_{\rho},p_{z})$ using cylindrical coordinates.
The electron is exposed to the potential $V$ and driven by a laser field linearly polarized along $\hat z$. The probability to measure a photo electron with momentum $P_{z}$, ejected along the laser polarization axis $\hat z$, is given as a two-dimensional integral over initial phase-space variables (denoted with a prime),
\begin{equation}\label{eq:spectrum} 
{\cal P}(P_{z}) = \int\!\!\!\int\d\phi' \d p_{\!\rho}'\,w(\phi', p_{\!\rho}')\,\delta\left(P_{z}-p_{z}(\phi',p_{\!\rho}')\right)\,,
\end{equation} 
where $\phi' \equiv \omega t'$ is the phase of the laser at the time when the electron tunnels and $p_{\!\rho}'$ is the initial momentum perpendicular to the tunneling direction $\hat z$. 
The weight $w(\phi', p_{\!\rho}')$ accounts for the tunnel probability and Jacobian factors. 
The relevant dynamical object in a classical dynamical theory is the deflection
function $p_{z}(\phi',p_{\!\rho}')$ which relates final variables to the initial conditions of the trajectory \cite{ro98}. 
\Fig{fig:contours123} shows $p_{z}$ as a function of the initial phase $\phi'$ (or rather the corresponding vector potential $A'=A\sin\phi'$ with $A\equiv F/\omega$) and the initial transverse momentum $\prhoi$.
One can see that 
$p_{z}$ develops
``finger-like'' structures with increasing time. They emerge first in the second laser period and
with each period an additional finger appears.
 These fingers are due to head-on collisions, responsible for the well-known high-energy phenomena such as HHG and ATI. Also, these regions are characterized by chaotic dynamics \cite{saro99}, very sensitive
 to initial conditions. 
Responsible for the distinct peaks in ${\cal P}(P_{z})$ in \fig{fig:contours123}d at low energies, however, are \emph{not} the fingers, but the crossings of contour lines above the fingers in
\fig{fig:contours123}b,c. They represent saddle points in
$p_{z}(\phi',p_{\!\rho}')$ with 
\begin{align}
&\partial p_{z}/\partial \phi'= 0,\quad \partial
p_{z}/\partial p_{\!\rho}'=0,
\notag\\ \label{eq:saddle}
&\left(\partial^{2} p_{z}/\partial\phi'{}^{2}\right)
\left(\partial^{2} p_{z}/\partial p_{\!\rho}'^{2}\right)<0. 
\end{align}
Such two-dimensional saddles are known to produce peaks since they represent integrable singularities in the spectrum \footnote{The situation is similar to the logarithmic singularities of the density of states in a two-dimensional crystal: L.~V. Hove, Phys. Rev. {\bf 89}, 1189 (1953).}. With each additional laser period (for the first three we show contour plots and for the first four we show the spectra in \fig{fig:contours123}) a new peak appears.
This establishes, that the LES actually consists of a series of peaks converging towards threshold $P_{z}=0$. 

\begin{figure}[b]
\begin{center}
\includegraphics[width=0.38\columnwidth]{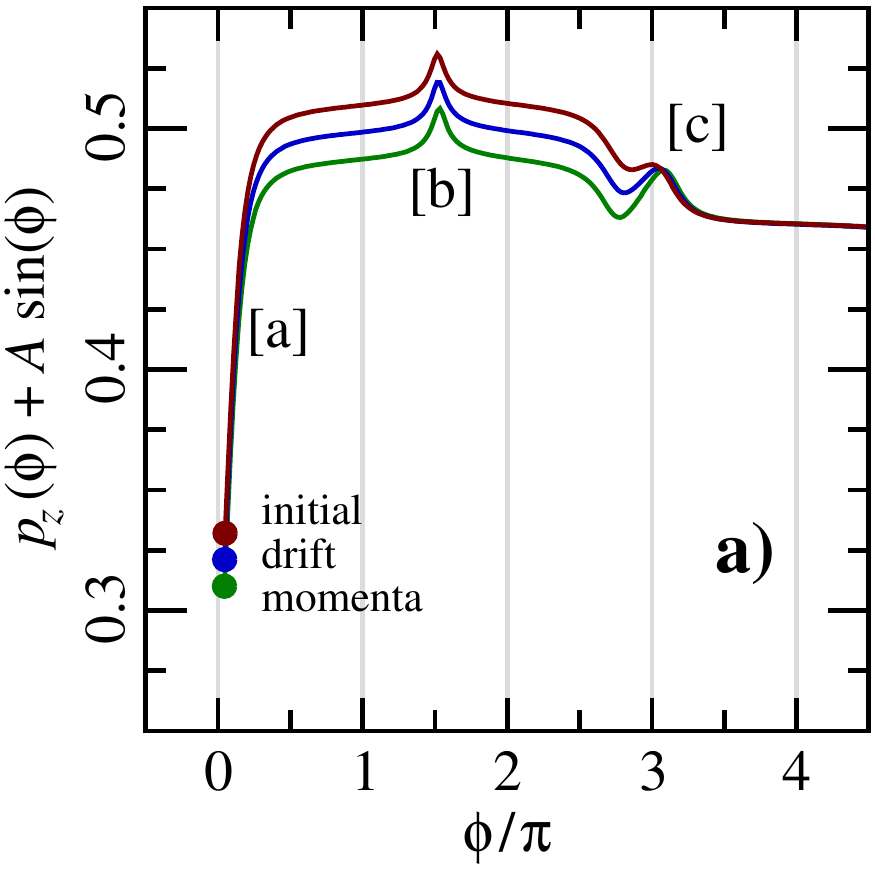}
\includegraphics[width=0.6\columnwidth,angle=0]{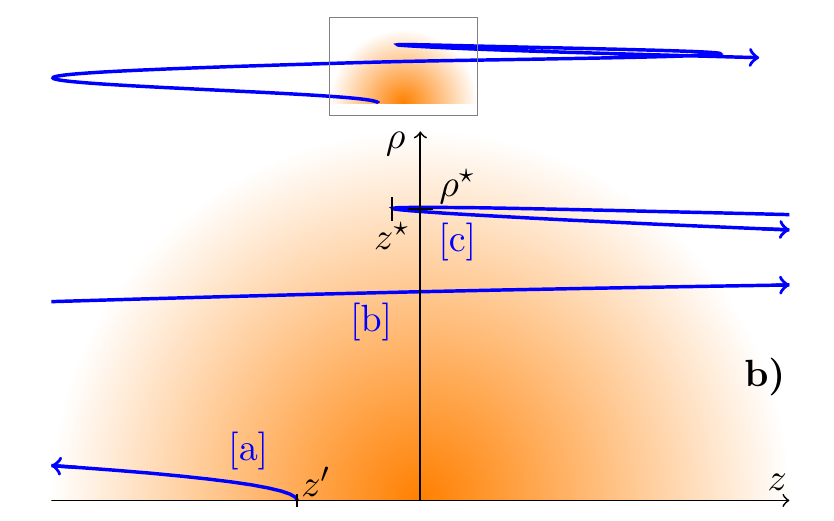}
\caption{(Color online) Time-dependent drift momentum $p_{z}(\phi)+A\sin(\phi)$ for three trajectories (left panel) with different initial drift momenta showing the effect of the Coulomb potential after release \coll{a} and the bunching during the soft recollision \coll{c}.
Sketch of the rescattering trajectory (right) in a Coulomb potential (orange-shaded area). Full trajectory (upper panel) and details (lower) of the three interactions events:
\coll{a} emission at $\phi\approx0$, \coll{b} effectless passing at $\phi\approx3\pi/2$, and \coll{c} soft recollison at $\phi\approx3\pi$.}
\label{fig:bunchingandtrajectory}
\end{center}
\end{figure}

Figure \ref{fig:bunchingandtrajectory}a illustrates that these peaks are due to an energy-bunching mechanism of neighboring trajectories: The three trajectories shown, start at similar but different phases $\phi'$ with corresponding drift momenta $A\sin\phi'$.
However, they carry the same momentum after the ``soft recollision'' with the nucleus in the second laser period at $\phic\approx3\pi$. The trajectory shown in \fig{fig:bunchingandtrajectory}b (corresponding to the central trajectory of \fig{fig:bunchingandtrajectory}a) reveals, that it is a soft recollision with the nucleus which leads to the energy bunching: The
electron ``misses'' the nucleus ($\rhoc>0$) and recollides by virtue of the laser force which turns the electron around at $\zc$ with $|\zc|\ll\zquiv\equiv F/\omega^{2}$. Hence, this new type of a recollision
is quite different from the elastic reflection off the potential with finite momentum $\pc$ as in the head-on collisions in ATI or HHG. How do these soft recollisions provide a series of peaks?

With the characteristics of the soft recollision (all related quantities are denoted with a star) as observed 
\begin{equation}\label{eq:softcondition}
\zc\equiv z(\phic) \sim 0\:\mbox{ and }\: \pc\equiv p_{z}(\phic) = 0\,,
\end{equation}
this is easy to see using strong-field trajectories
\begin{subequations}
\label{eq:sftrajectories}
\begin{align}
\label{eq:sftrajectories-z}
z(\phi) &= \zquiv\left([\phi'{+}\phiV]\phi + \cos\phi{-}1\right)
\\
\label{eq:sftrajectories-p}
p_{z}(\phi) &= A[\phi'{+}\phiV] - A\sin\phi\,.
\end{align}
\end{subequations}
where we have linearized the solutions of Hamilton's equations for $H_{0}$ in $\phi'$ since tunneling occurs near the maximum of the field $F\cos\phi' \sim F$, i.\,e., $\phi'\ll 1$. Moreover,
$\phiV=\Delta p/A$ accounts for an overall $\Delta p$ offset of the drift momentum due to the potential.
As can be seen in \fig{fig:bunchingandtrajectory}a this offset does not dependent on $\phi'$. 

From $p_{z}(\phic)=0$ we get with \eq{eq:sftrajectories-p} immediately $\phic = m\pi +(-1)^{m}[\phi'{+}\phiV]$.
A little thought reveals that 
only odd integers $m = 2n+1$ yield non-trivial solutions. 
Requiring that $\zc=0$ with the recollision location
\begin{equation}\label{coll-zc}
z(\phic) = \zquiv[(2n{+}1)\pi[\phi'{+}\phiV]-2]=0
\end{equation}
gives the initial phases $\phi'_{n}$ and in turn the drift momenta $p_{n} =A[\phi'_n{+}\phiV]$, cf.\ \eq{eq:sftrajectories-p}, for the soft recollisions
\begin{equation}\label{eq:peaks}
p_n = \frac{F/\omega}{(n{+}1/2)\pi} \,.
\end{equation}
From this equation
we expect a series of photo-electron momentum peaks. This is indeed confirmed by the spectra shown in \fig{fig:contours123}d, where peaks appear cycle by cycle.
How do these peaks emerge? 

\begin{figure}[t]
\begin{center}
\includegraphics[width=\columnwidth]{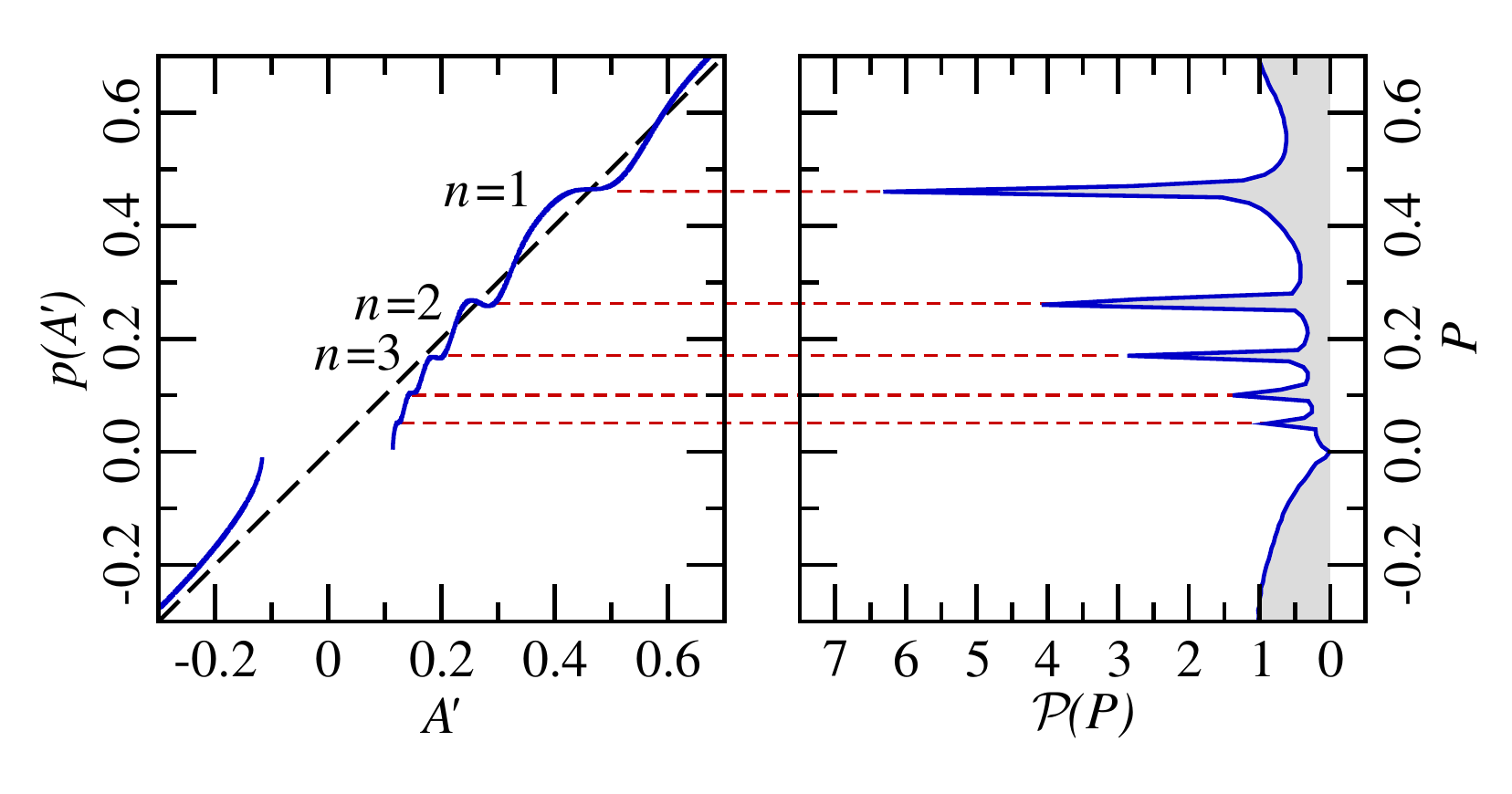}
\caption{(Color online) Deflection function (left) and corresponding photo-electron spectrum with finite resolution
(right) for the Gaussian potential \eq{eq:gausspot}, laser parameter as in \fig{fig:contours123}. The missing interval in the deflection function represents initial conditions which lead to trapped trajectories. The dashed line corresponds to the strong field drift momentum $p(\phi') = A\sin\phi'$.}
\label{fig:gausspot} 
\end{center}
\end{figure}

From our analysis so far, one-dimensional (1D) dynamics with some potential (short or long range) should be sufficient to explain the underlying mechanism.
To this end we consider the 1D Hamiltonian $H=p^{2}/2+xF\cos(\omega t)+V_{s}(x)$, with position $x$, momentum $p$ and the range $s$ of the potential.
Starting always at the origin $x=0$, but with different phases $\phi'$, the electron is propagated until $|x|\gg s$ and the drift momentum $p(\phi){+}A\sin(\phi)$ is constant. 
The deflection function $p(\phi')$ along with the corresponding photo-electron spectrum
\begin{subequations}\label{eq:model}\begin{align}
{\cal P}(P) & = \frac 1{2\pi}\int\d \phi'\,\delta\left(P-p(\phi')\right) 
\label{eq:model1}
\\ 
& = \frac 1{2\pi}\sum_{i}\left|\frac{\d p}{\d\phi'}\right|^{-1}_{p(\phi'_{i}) = P}\,,
\label{eq:model2}
\end{align}\end{subequations}
is shown in \fig{fig:gausspot} for the short-range Gaussian potential
\begin{equation}\label{eq:gausspot}
V_{s}(x) = -\exp\left(-(x/s)^{2}\right)/s\,.
\end{equation}
By integrating the force due to the potential $V_{s}$ the deflection function 
can be written in the form
\begin{subequations}\begin{align}\label{momentum}
p(\phi') &= A\phi' +\delta p(\phi')\\ 
\label{eq:impact}
\delta p(\phi') &= - \frac 1\omega \int_{\phi'}^{\infty}\!\!\mathrm d\phi
\left.\frac{\d V_{s}}{\d x}\right|_{x=x(\phi)} ,
\end{align}\end{subequations}
where $A\phi'$ (shown as a dashed line in \fig{fig:gausspot}a) represents the contribution from the laser field without the potential. The second term
$\delta p$ represents the impact from the external potential 
which leads to modulations in $p(\phi')$.
Whether the modulations are really visible as pronounced peaks in the spectrum depends on the strength of the impact $\delta p(\phi')$ in \eq{eq:impact}.

Peaks occur in the first place if $dp/d\phi'= 0$, cf.\ \eqref{eq:model2}. Physically, this means that 
the change in the impact strength $d\delta p/d\phi'$ must exactly compensate the change in the drift momemtum, which is simply $A$. A weak impact leads only to a marginal decrease of the slope of $p(\phi')$ giving rise to a shallow hump in the spectrum. On the other hand a strong impact $\delta p$ will overcompensate
the change of the drift momentum leading to a negative slope for some $\phi'$ accompanied by two extrema.
This is indeed the case in \fig{fig:gausspot} for the higher recollisions $n>1$. 
The cross-over between weak and strong impacts is determined by
$d^{2}p/d\phi'^{2}= 0$, which we take as a measure for a potential to produce pronounced peaks.
These two conditions allow us to determine the initial phase $\phi'$ and
the strength parameter $s$ of the potential for producing pronounced peaks through soft collisions as
we will show now analytically.
To this end we consider the integral, cf.\ Eq.\,\eqref{eq:impact}
\begin{subequations}\label{eq:impactintegral}
\begin{equation}
\delta p(\phi') = - \frac 1\omega \int_{-\infty}^{+\infty}\!\!\mathrm d\phi
\left.\frac{\d V_{s}}{\d x}\right|_{x=x(\phi)}
\end{equation}
for a soft-recollision trajectory
\begin{equation}
x(\phi)=\xc(\phi')+ \frac{\xquiv}{2}(\phi{-}\phic)^{2}.
\end{equation}
\end{subequations}
It is defined by the quadratic dependence of a strong-field trajectory around the recollision phase $\phic$.
One can extend this behaviour to $\phi\to\pm\infty$, since the force in Eq.\,(\ref{eq:impactintegral}a) vanishes for large $|\phi|$.
Note that the impact $\delta p$ depends through the recollision point $\xc$ on the initial phase $\phi'$. 

Fullfilling the condition $d^{2}p/d\phi'^{2}= 0$ can be cast into the form $f_{2}(\cc) = 0$ for the ratio $\cc\equiv\xc/s$ and the function 
\begin{subequations}\label{eq:generalf}\begin{align}
\label{eq:generalf1}
f_{j}(c^{\star}) &\equiv -\frac{k_{j}}{\omega}\int_{-\infty}^{+\infty}
\!\!\!\mathrm{d}\phi\left.\frac{d^{j+1}V_{1}(x)}{dx^{j+1}}\right|_{x=c^{\star}+\phi^{2}/2}
\\
\label{eq:generalf2}
k_{j} &\equiv\left[(2n{+}1)\pi\right]^{j}\xquiv^{j-1/2}\big/s^{j+3/2}.
\end{align}\end{subequations}
This follows directly from Eq.\,\eqref{eq:impactintegral} by using the chain rule and an appropriate rescaling of the integration variable. 
Note, that the integral in Eq.\,\eqref{eq:generalf1} does only depend on the \emph{shape} of the potential, but not on any specific parameters of the problem $s$, $F$, or $\omega$. 
Hence, the value $\cc$ which solves $f_{2}(\cc) = 0$ is a general constant which assumes the value $\cc = -0.319$ for the Gaussian potential \eqref{eq:gausspot}. 
The first condition $dp/d\phi'= 0$ reads with the definition \eq{eq:generalf} simply
$A+f_{1}(\cc)=0$ which can be recast into a form that determines $s$ as a function of the laser parameters $F$, $\omega$ and the order of the recollision $n$,
\begin{equation}\label{eq:svalue}
s = \frac{[2(2n{+}1)\pi f_{1}(\cc)]^{2/5}}{(F\omega^{2})^{1/5}}.
\end{equation}
This allows us to determine quantitatively the scale $s$ and through the relation $\xc/s = \cc$ also the point of the recollision $\xc$ at which the deflection function has a zero-slope inflection point.
Our quasi-analytical determination of the soft-collision parameters $s$ and $\xc$ is remarkably accurate as the comparison with the numerical exact values from the soft colliding trajectory propagated under $H$ reveals in Table I. 
There, we also list the corresponding values for the 1D soft-core Coulomb potential
\begin{equation}\label{eq:softcore}
V_{s}(x) = -1/\left(x^{2}{+}s^{2}\right)^{1/2},
\end{equation}
for which one obtains $\cc=-0.264$.
\begin{table}
\begin{ruledtabular}
\caption{Soft-recollision parameters (in a.u.) for the Gaussian \eqref{eq:gausspot} and the soft-core \eqref{eq:softcore} potential as obtained from Eqs.\,\eqref{eq:svalue} and \eqref{eq:rhovalue}, respectively.
For comparison full numerical results for propagation until $t=2T$ and $t=3T$, respectively, are shown in italics for the laser parameters of \fig{fig:contours123}.}
\begin{tabular}{rcccc|rc}
&\multicolumn{2}{c}{1D Gaussian} &\multicolumn{2}{c}{1D soft-core} & \multicolumn{2}{c}{3D Coulomb}\\
& $n=1$ & $n=2$ & $n=1$ & $n=2$ && $n=1$ \\\hline
$s$ & 32.7 &  40.2 &  24.6 &  30.2 & $\rhoc$ & 23.6 \\ [-2pt]
& {\it 31.9} & {\it 38.4} & {\it 23.9} & {\it 29.4} & & {\it 22.2} \\
$-\xc$ & 10.5 &  12.8 &  6.5 &  8.0 & $-\zc$ & 10.9 \\ [-2pt]
& {\it 10.2}& {\it 12.6}& {\it 6.4}& {\it 8.1} & & {\it 12.4} \\
\end{tabular}
\end{ruledtabular}
\end{table}%

In fact, the 3D physical case discussed in the beginning can be mapped onto the 1D soft-core potential since $\rho$ is very slowly varying across the soft collision 
(see \fig{fig:bunchingandtrajectory}b) and can be effectively treated as a parameter, i.\,e., we take at the soft collision $\rhoc=s$ in the soft-core potential \eqref{eq:softcore}. In order to fullfill the saddle-point 
conditions \eqref{eq:saddle} we only have
to exchange $d^{2}p/d\phi'^{2}=0$ from our 1D treatment with $\partial p_{z}/\partial \rho =0$.
The latter reads 
\begin{equation}\label{eq:integralford}
\int_{-\infty}^{+\infty}\!\!\mathrm{d}\phi\left[2\frac{dV}{dz}+z\frac{d^{2}V}{dz^{2}}\right]_{z=\Cc+\phi^{2}/2}=0,
\end{equation}
and can be expressed with integrals from \eq{eq:generalf} producing an equation only dependent on
 $\Cc = \zc/\rhoc$. 
A similar procedure as described above for the 1D case yields $\Cc=-0.462$ and ultimately
\begin{equation}\label{eq:rhovalue}
\rhoc= 2.90/\left(F\omega^{2}\right)^{1/5}
\end{equation}
for the first recollision ($n=1$) in very good agreement with the numerical values, see Table I.

In summary, we have identified a soft-recollision mechanism which induces energy bunching for low-energy photo electrons along the laser polarization. The bunching occurs since electrons with initially different drift momenta
can aquire impacts through soft recollisions which exactly counterbalance the initial differences
leading to a series of photo-electron peaks with relative positions 
$p_{n}/p_{n-1}=(2n{-}1)/(2n{+}1)$ for $n{>}1$.
This is a universal result which does neither depend on the dimensionality of the potential (one degree of freedom is enough) nor on the character of the potential (short or long range) or 
the laser intensity and frequency.
It does, however, require a quiver amplitude $\zquiv$ much larger than the range of the potential $s$.
This is necessary to provide well defined impacts $\delta p$ by the potential when the mainly laser driven electron trajectory passes the potential. 

The absolute positions $p_{n}$ of the peaks are slightly dependent on the potential and the laser pulse and will also be influenced by focal-volume averaging and the pulse envelope. 
This applies in particular to the higher-order peaks very close to threshold where also additional dynamical effects may mask the soft recollision peaks.
However, since each peak $p_{n}$ is generated in a successively later laser period $n{+}1$, one can in principle control the number of peaks by varying the total number of cycles in the laser pulse \footnote{A. K\"astner, U. Saalmann, and J.\,M. Rost, in preparation}. The phenomenon is essentially classical because the potential perturbs the strong field dynamics only marginally. The latter contains only up to quadratic operators
(depending on the length or velocity gauge). Owing to the Ehrenfest theorem the quantum evolution 
can therefore be described equivalently by classical mechanics.

\end{document}